\documentclass[aps,pre,reprint]{revtex4-2}

\draft 
\usepackage{amssymb,amsmath}
\usepackage{graphicx}
\usepackage{dcolumn}
\usepackage{bm}
\usepackage{color}
\usepackage{float}
\usepackage{hyperref}
\usepackage[version=4]{mhchem}
\usepackage{relsize}
\usepackage{comment}
\usepackage{media9}

\newcommand{\Rey}{{\mathcal Re}}

\begin{document}

\title{Analogue and Physical Reservoir Computing Using Water Waves}

\author{Ivan S.~Maksymov}
\email{imaksymov@csu.edu.au}
\affiliation{Artificial Intelligence and Cyber Futures Institute, Charles Sturt University, Bathurst, NSW 2795, Australia\looseness=-1}


\begin{abstract}
More than 3.5\,billion people live in rural areas, where water and water energy resources play an important role in ensuring sustainable and productive rural economies. This article reviews and critically analyses the recent advances in the field of analogue and reservoir computing that have been driven by unique physical properties and energy of water waves. It also demonstrates that analogue and reservoir computing hold the potential to bring artificial intelligence closer to people living outside large cities, thus enabling them to enjoy the benefits of novel technologies that already work in large cities but are not readily available and suitable for regional communities.
\end{abstract}

\maketitle 

\section{Introduction\label{sec:1}}
\begin{figure}[t]
 \includegraphics[width=0.49\textwidth]{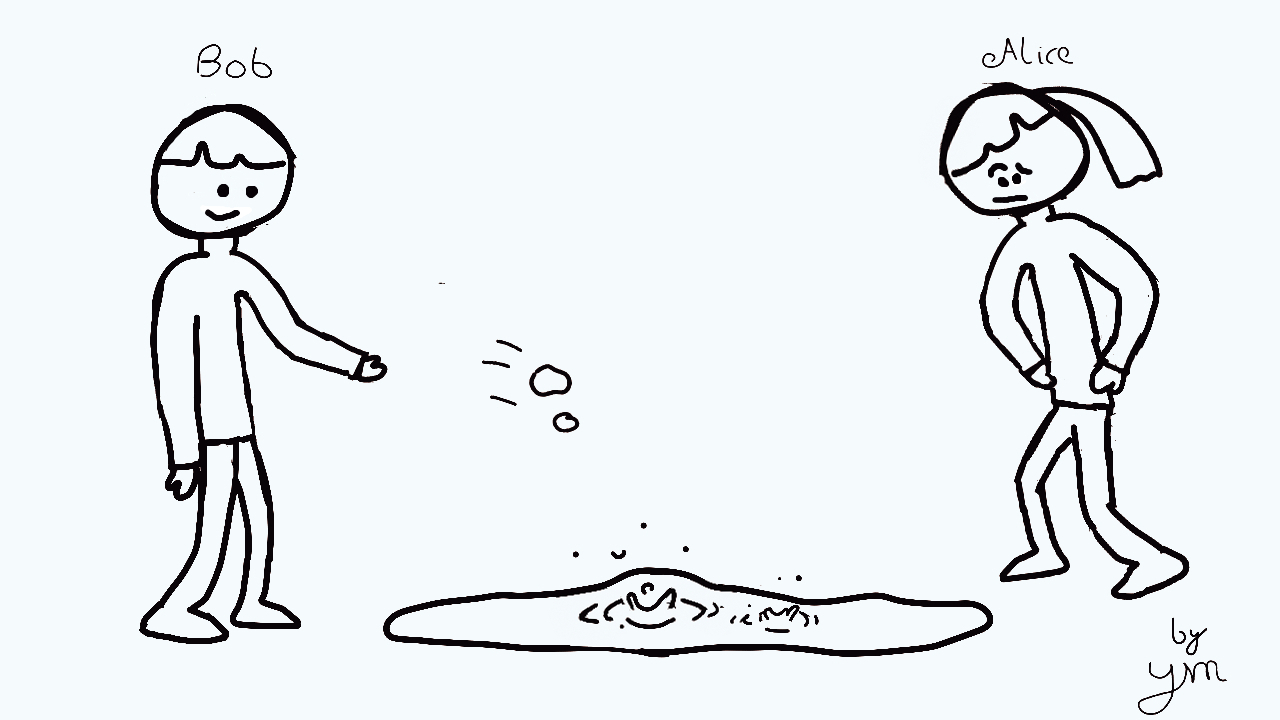}
 \caption{Illustration of the game played by Alice and Bob that can be used to explain the operation of reservoir computing systems.\label{Fig1}}
\end{figure}

Ever since computers were invented, one of the main questions raised by experts and the general public has been whether machines might learn to improve themselves automatically similarly to a biological brain \cite{Puc67, Mit97, Sch21}. To put this question into perspective, let us imagine two kids, Alice and Bob, playing at the edge of a pond (Figure~\ref{Fig1}). Bob chooses the stones randomly and drops them into water. He believes that Alice cannot predict the size of the falling stones. However, watching and analysing the wave patterns created by the stones of different size, Alice learns to predict the size of the stones that Bob is going to drop next.

In this imaginary game, the actions of Bob may be regarded as the response of a dynamical system that exhibits a chaotic behaviour \cite{Wei93, Sma05, Shu17} but Alice's ability to learn and predict Bob's actions in the future may represent the operation of certain machine learning (ML) algorithms \cite{Mit97, Hay98, Mar20} designed to forecast the time evolution of real-life dynamical systems such as stock markets, autonomous vehicles, the behaviour of living organisms and variation of the Earth's climate \cite{Wei93, Mar04, Sma05, Shu17}.

Of course, even a highly skilled human cannot produce detailed long-term forecasts of global financial markets and climate change. Nevertheless, learning from a biological brain, we can create ML systems that can mimic some functions of the brain and process large amount of certain classes of data more efficiently than a human \cite{Hay98, Shu17}. However, complex nonlinear properties of many natural and artificial dynamical systems considerably complicate the task of computerised prediction and, therefore, force ML algorithms to rely on longer observation times, thus demanding substantial computational resources for processing big data sets.

While high-performance computational resources needed to process big data sets are readily accessible to large companies, governmental organisations and research and education sector, the options available to small businesses and individuals are mostly limited to the use of third-party and open-source artificial intelligence (AI) software run on a cloud-based infrastructure or personal computers. Yet, even these limited opportunities may not be available to people living in the developing countries and in rural and remote communities \cite{Ganna}. Subsequently, there is a need for free and easy-to-use ML systems that can operate independently of mainstream information technology services.

\begin{figure*}[t]
 \includegraphics[width=0.69\textwidth]{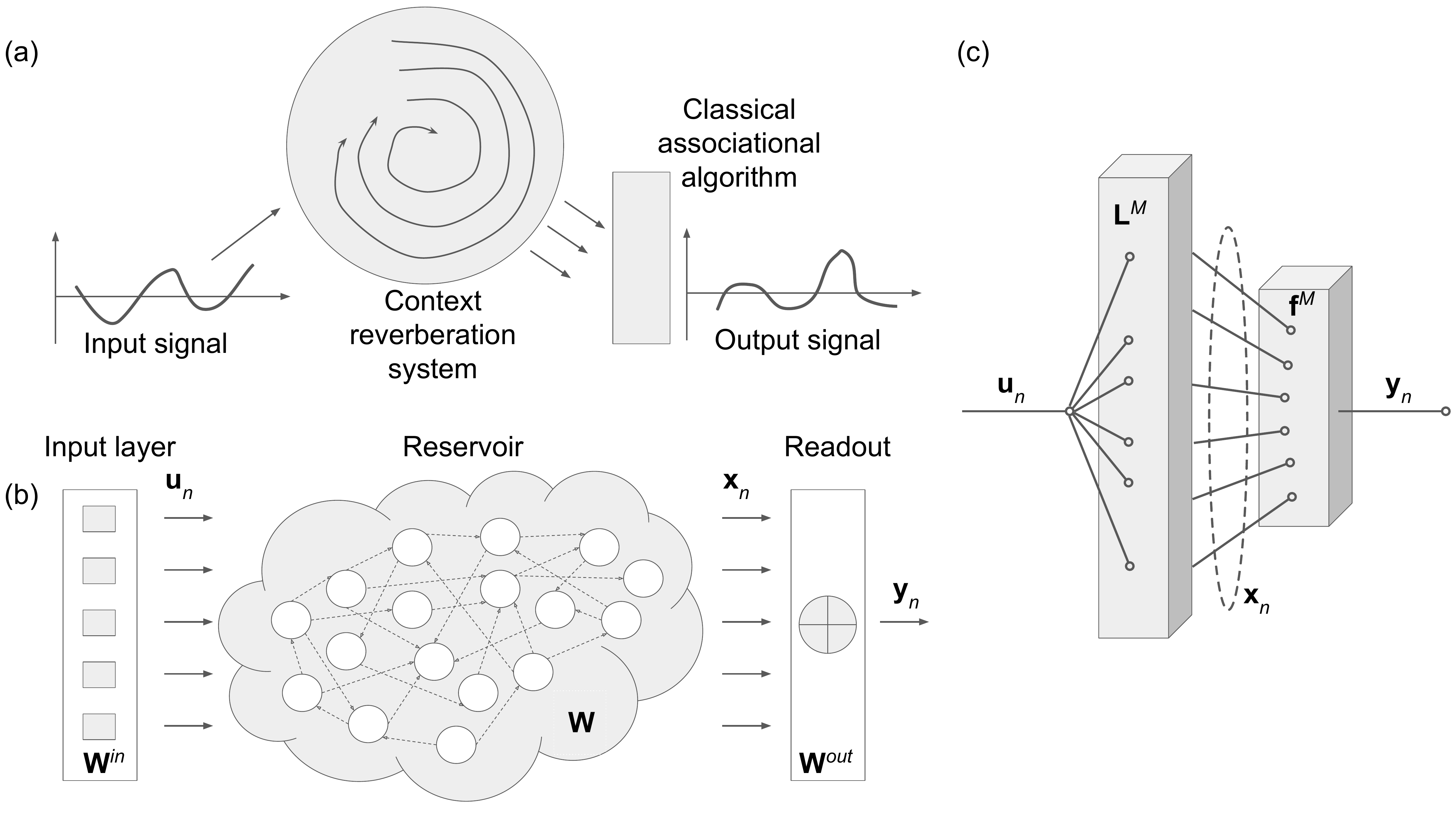}
 \caption{{\bf (a)}~Illustration of Kirby's concept of the ``context'' of compressed (reverberated) input histories processed by a classical associational algorithm (linear readout). {\bf (b)}~Block diagram of an ESN-based RC algorithm. The reservoir is a network of interconnected artificial neurons that produces a vector of neural activations ${\bf x}_n$ using a data set of input values ${\bf u}_n$. Only the linear readout is trained to produce an output ${\bf y}_n$. {\bf (c)}~Architecture of an LSM. A data set of input values ${\bf u}_n$ is used as the input into the abstract liquid (${\bf L}^M$) to create liquid states ${\bf x}_n$ that are then employed by a linear readout to generate an output ${\bf y}_n$. Note a conceptual similarity between the three algorithms.\label{Fig2}}
\end{figure*}

The game depicted in Figure~\ref{Fig1} illustrates one potential solution of the problem of accessible AI---reservoir computing (RC) \cite{Luk09, Ver07, Gau21, Nak21}. Based on the paradigms of ``context'' of compressed reverberated input histories \cite{Kir91}, Liquid State Machines (LSMs) \cite{Maa02, Maa04} and Echo State Networks (ESNs) \cite{Jae05, Luk09} (Figure~\ref{Fig2}) and several other relevant artificial neural network architectures \cite{Sch91}, RC is an emergent class of ML algorithms designed to forecast the response of nonlinear dynamical systems that exhibit chaotic behaviour \cite{Jae05, Pat17, Pat18, Cha20, Gau21}. An RC system can perform such complex tasks due to its special structure (Figure~\ref{Fig2}) consisting of a fast output layer and a slow artificial neural network called the reservoir \cite{Jae05} (the notion of abstract liquid is used instead of the term ``reservoir'' in the framework of LSM \cite{Maa02}).

Typically, the reservoir is a large network of randomly connected artificial neurons that communicate one with another and also receive time-dependent input from external data sources. The role of this network consists in converting a time varying input into a spatio-temporal pattern of neural activations (in the framework of LSM neural activations are called liquid states). The resulting neural activations are then transformed into practical forecasts using the readout layer. Hence, returning to the illustration in Figure~\ref{Fig1}, we can say that an RC system mimics a stone that is dropped into water and that produces ripples on the water surface. The falling stone plays the role of an external data input since its motion is transformed into a spatio-temporal pattern of the ripples.

Thus, since a reservoir trained on a particular data set can be used to produce many useful outputs, RC systems can process data with a higher speed than many competitive ML techniques while using standard computational resources \cite{Luk12, Bol21}. Yet, even though the operating principle of an RC system does not necessarily replicate essential functions of a biological brain \cite{Maa02, Luk09}, an analogy may be drawn between a reservoir and some biological brains, including the brains of insects that have just 200,000 neurons, compared with billions neurons in a mammalian brain, but enable them to perform complex tasks while they navigate and search for food \cite{Raj21}. Demonstrations of single-neuron reservoir computers \cite{App11} strongly speak in favour of this statement. The concept of deep learning also applies to RC systems that consist of several interconnected reservoirs and that can learn from large amounts of data using standard computational resources \cite{Gal17}

In the focus of this review article is another intriguing property of RC systems---their ability to operate using a physical reservoir \cite{Nak21, Tan19, Nak20, Cuc22}. This property holds a potential to make reservoir computers accessible to millions of users for a low cost. While ESN and LSM are, essentially, computer programs, it has been suggested that their algorithms can be implemented using certain nonlinear physical systems such as different types of waves \cite{Nak21, Tan19, Nak20, Hug19}. At the conceptual model level, this means that the equations that constitute the backbone of the ESN and LSM algorithms (see Sect.~\ref{sec:ESN}) become replaced with the actual response signal of a nonlinear physical system \cite{Hug19, Mak21_ESN}. Similarly to analog computers that can solve certain problems more efficiently than digital computers \cite{Hyn70, Cow05, Sor15, Zan18}, physical RC systems may be more energy and computationally efficient than an algorithmic RC in practical situations, where the relationship between time-dependent physical quantities that needs to be predicted can be correctly represented by the dynamics of the physical system. In practice, this condition can be satisfied by many physical systems, including spintronic devices \cite{Fur18, Rio19, Wat20}, quantum ensembles \cite{Fuj17}, electronic devices \cite{Cow05, Pen18, Nak21}, photonic systems \cite{Nak21, Sor20, Zen20, Sil21, Pan21, Che20_review, Raf20} and mechanical devices \cite{Nak21, Cou17}. Some of these physical systems can also be used to understand and predict the behaviour of financial markets \cite{Lil00, Pet13, Nas14}. 

In particular, in this article we discuss the implementations of the concept of LSM using liquids. Although the creators of LSM employed the analogy between a reservoir and an abstract liquid \cite{Maa02}, it has been demonstrated that physical liquids inherently possess nonlinear dynamical properties needed for the creation of a reservoir \cite{Fer03, Jon07, Nak15, Nak20, Got21, Mar20, Mak21_ESN, Mat22}. Thus, we critically review the results presented in these works, also discussing the potential practical applications of physical LSM-based RC systems.

Before we commence our discussion, we note that a number of previous articles already review the theoretical foundations and practical applications of physical RC systems \cite{Luk09, Nak20, Nak21, Tan19, Cuc22, Che20_review, Bal18, All23}. Although those works are focused mostly on physical RC systems based on electronic, photonic and magnonic devices, some of them also discuss liquid-state systems. Therefore, to reduce the overlap with the previously published surveys, in this current article we generally review the seminal works of liquid-state physical RC systems and then discuss the results produced in the recent years.     

\section{Algorithmic Reservoir Computing}
\subsection{Context of Reverberated Input Histories}
Although the concept of RC was established as a result of a series of pioneering works published in the 2000s \cite{Jae01, Maa02, Maa03, Jae04, Jae05}, many literature sources overlook an earlier work by K.~G.~Kirby \cite{Kir91}, where what is currently called the computational reservoir was originally introduced as a context of compressed (reverberated) input histories (Figure~\ref{Fig2}a). Kirby introduced an artificial neural network structure designed for efficient learning of spatio-temporal dynamics, where the network architecture was split into two subproblems: the formation of the context and the classification of context by an association algorithm. He suggested that the first subproblem could be resolved representing the neural network by a nonlinear dynamical system such as a low-connectivity random network or a continuous reaction-diffusion model. Furthermore, he demonstrated that using the so-created neural network enables resolving the second subproblem using a linear learning algorithm that can be chosen from a wide range of well-established techniques \cite{Vee95, Mit97, Sma05, Gal07}. ESN, LSM and similar artificial neural network architectures follow virtually the same approach (Figure~\ref{Fig2}b,~c).

\subsection{Echo State Network\label{sec:ESN}}
The mathematical nonlinear dynamical system that underpins the operation of the artificial neural network of ESN is governed by the following update equation \cite{Jae05, Luk09, Luk12}:
\begin{eqnarray}
  {\bf x}_{n} = (1-\alpha){\bf x}_{n-1}+
  \alpha\tanh({\bf W}^{in}{\bf u}_{n}+{\bf W}{\bf x}_{n-1})\,,
  \label{eq:RC1}
\end{eqnarray}
where $n$ is the index denoting entries corresponding to equally-spaced discrete time instances $t_n$, ${\bf u}_n$ is the vector of $N_u$ input values, ${\bf x}_n$ is a vector of $N_x$ neural activations of the reservoir, the operator $\tanh(\cdot)$ applied element-wise to its arguments is a typical sigmoid activation function used in the nonlinear model of a neuron \cite{Hay98}, ${\bf W}^{in}$ is the input matrix consisting of $N_x \times N_u$ elements, ${\bf W}$ is the recurrent weight matrix containing $N_x \times N_x$ elements and $\alpha \in (0, 1]$ is the leaking rate that controls the update speed of the reservoir's temporal dynamics.

To train the linear readout of ESN one calculates the output weights ${\bf W}^{out}$ by solving a system of linear equations ${\bf Y}^{target} = {\bf W}^{out}{\bf X}$, where the state matrix ${\bf X}$ and the target matrix ${\bf Y}^{target}$ are constructed using, respectively, ${\bf x}_n$ and the vector of target outputs ${\bf y}_n^{target}$ as columns for each time instant $t_n$. The solution is often obtained in the form ${\bf W}^{out} = {\bf Y}^{target} {\bf X^\top} ({\bf X}{\bf X^\top} + \beta {\bf I})^{-1}$, where ${\bf I}$ is the identity matrix, $\beta=10^{-8}$ is a regularisation coefficient and ${\bf X^\top}$ is the transpose of ${\bf X}$ \cite{Luk12}. Then, one uses the trained ESN, solves Eq.~(\ref{eq:RC1}) for new input data ${\bf u}_n$ and computes the output vector ${\bf y}_n={\bf W}^{out}[1;{\bf u}_n;{\bf x}_n]$ using a constant bias and the concatenation $[{\bf u}_n;{\bf x}_n]$ \cite{Luk09, Luk12}.

\subsection{Liquid State Machine}
LSM \cite{Maa02} is another independent foundational concept of RC that was created simultaneously with ESN. However, unlike ESN, LSM was developed from a computational neuroscience point of view with a goal to study the behaviour of neural microcircuits \cite{Maa02, Maa04}. As a result of this difference, LSM employs more biologically realistic models of neurons and dynamic synaptic connections of the reservoir. In particular, the network of artificial neurons used in LSM often follows the topology of biological neural networks.

In the field of LSM, the reservoir is called the (abstract) liquid and an analogy is drawn between the excited neural states and ripples on the surface of a pond (which motivated the introduction of Figure~\ref{Fig1} and the discussion around it). However, at the other levels of its architecture LSM closely resembles ESN, including the use of sigmoid neural activation functions and linear readouts \cite{Maa02}. On the other hand, LSM also often employs more complex additional mechanisms not present in standard ESN such as averaging spike trains \cite{Cuc22}, which, however, increases the complexity of an LMS computer program compared with the one that implements the ESN algorithm. Subsequently, LSM is less widespread than ESN in the areas of applied physics, engineering and finance \cite{Nak21}. Nevertheless, since LSM mimics some operation principles of biological neurons, it can perform more complicated information processing tasks \cite{Ber04}.
\begin{figure}[t]
 \includegraphics[width=0.49\textwidth]{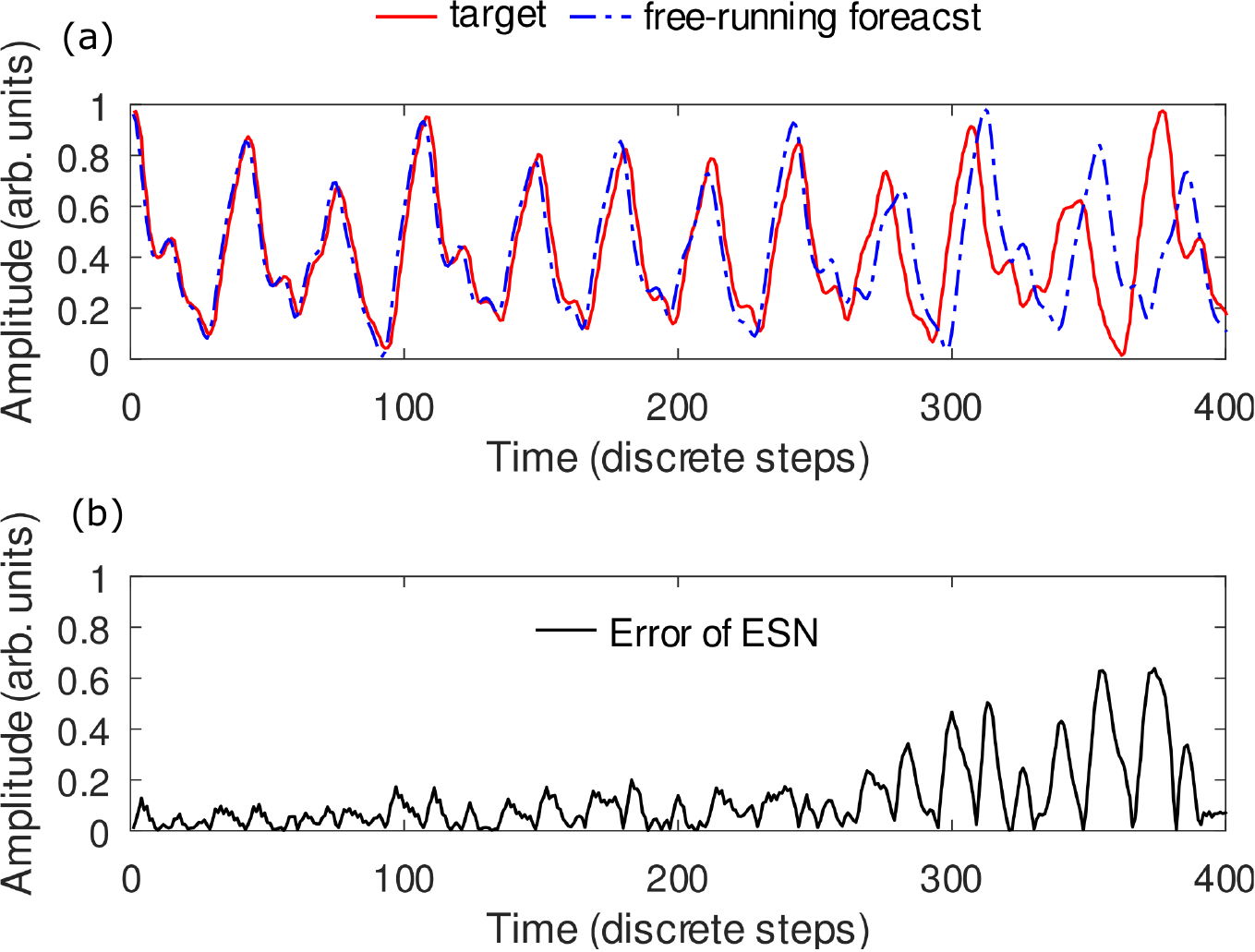}
 \caption{{\bf (a)}~Free-running forecast made by standard ESN compared with the target MGTS signal. {\bf (b)}~Modulus of the absolute error of the forecast made by ESN.\label{Fig3}}
\end{figure}

\subsection{Examples of Operation of ESN}
Since the standard ESN has been more often used in practice than LSM, in this section we demonstrate a typical scenario of operation of an RC system, where ESN is trained to forecast a chaotic nonlinear time series. Following a convention adopted in the literature on ESN and relevant ML algorithms \cite{Luk12}, as the first test time series we employ a chaotic Mackey-Glass time series (MGTS) \cite{Mac77}. Other chaotic time series of artificial \cite{Lor63, Ros76, Hen76, Ike79, Gau21, Mak21_ESN} and natural origin \cite{Gau21, Luk21} have also been employed to test ESN similarly to the result presented below.   

The MGTS is produced by the delay differential equation \cite{Mac77}
\begin{eqnarray}
  \dot{x}_{_{MG}}(t)
  &=&\beta_{_{MG}}\frac{x_{_{MG}}(\tau_{_{MG}}-t)}
      {1+x_{_{MG}}^{q}(\tau_{_{MG}}-t)}-\gamma_{_{MG}}x_{_{MG}}(t)\,,
  \label{eq:MG}
\end{eqnarray}
where one typically chooses $\tau_{_{MG}}=17$ and sets $q=10$, $\beta_{_{MG}}=0.2$ and $\gamma_{_{MG}}=0.1$ \cite{Luk12}. We first generate a sufficiently long MGTS data set and then split it into two parts. The first part is used to train the RC system but the second part is used as the target data. It is noteworthy that the RC system is presented with the MGTS data at the training stage only, but for making a free-running forecast it uses its own output as the input, i.e.~${\bf x}_{n}$ is calculated using Eq.~(\ref{eq:RC1}) with ${\bf u}_{n} = {\bf y}_{n-1}$. However, in this particular example we use the second part of the pre-generated MGTS as the target data that are needed to assess the accuracy of the forecast made by the RC system.

An example of a free-running forecast of MGTS made by standard ESN is presented in Figure~\ref{Fig3}a, where we used software described in Ref.~\cite{Luk12}. As we can see in Figure~\ref{Fig3}a, at the beginning of the modelled time interval ESN produces an accurate forecast, which is confirmed by Figure~\ref{Fig3}b, where the modulus of the absolute error of the forecats made by ESN is plotted. Then, we can see that the accuracy of the forecast deteriorates with time, though ESN remains able to correctly reproduce the essential dynamics of MGTS.

It is noteworthy that this particular result was obtained using a reservoir that contains 3000 neurons and is trained using a relatively short training time series and a typical parameter $\alpha=0.3$ in Eq.~(\ref{eq:RC1}) \cite{Jae01, Luk09, Bol21}. From the point of view of other standard ML algorithms, computational resources used by the so-configured ESN are surprisingly very readily affordable compared with the requirements of many competing ML techniques. Hence, the ability of ESN to produce a feasible forecast while using affordable computational resources makes it a best-in-class ML learning algorithm for prediction of the behaviour of highly nonlinear and chaotic dynamical systems \cite{Gau21}.
\begin{figure*}[t]
 \includegraphics[width=0.69\textwidth]{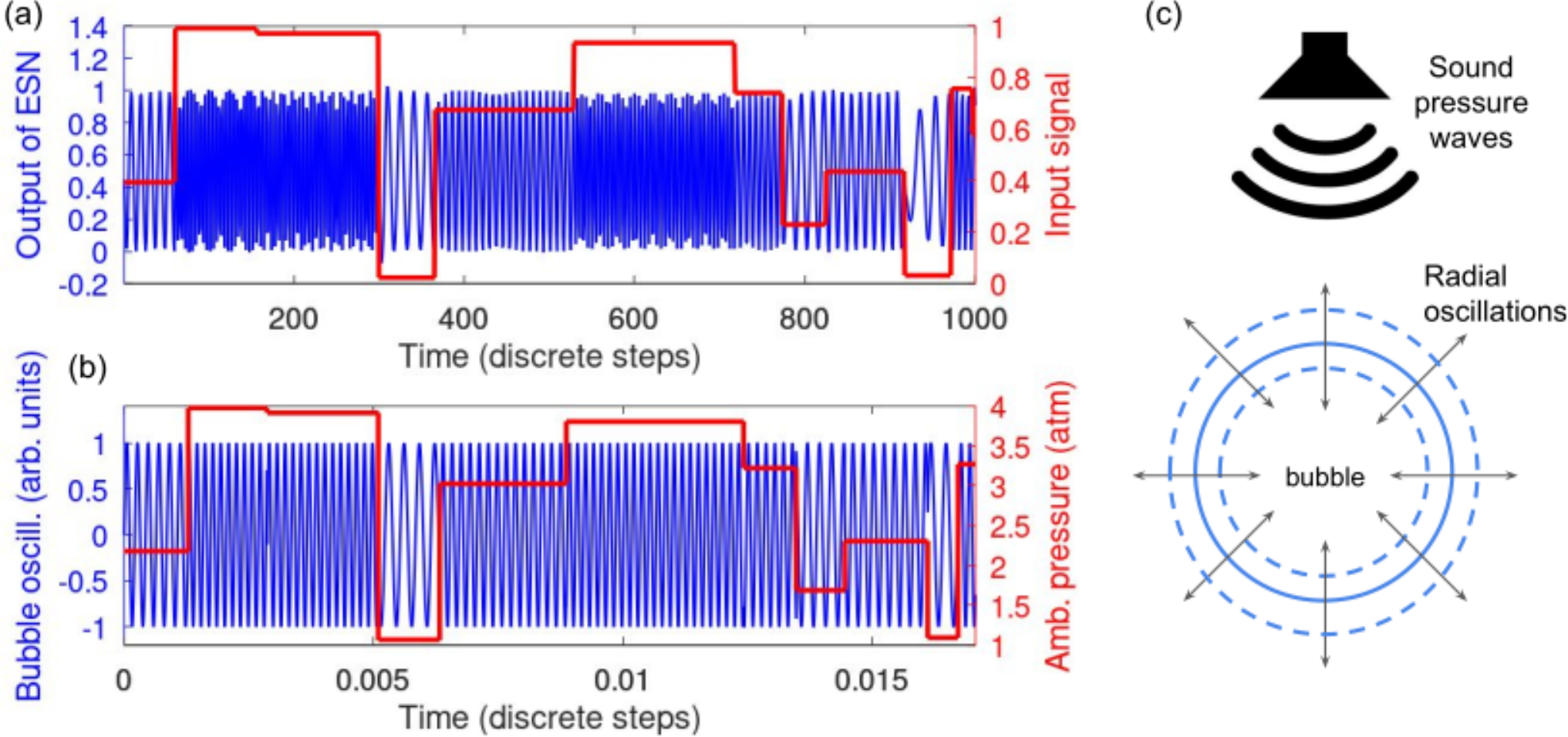}
 \caption{{\bf (a)}~Output of ESN trained to generate a sinusoidal signal with a frequency encoded as the input step pulse signal. {\bf (b)}~Response of a bubble in water that harmonically oscillates due to the changes in the ambient acoustic pressure (see panel~(c)) represented by the same step pulses as in panel~(a).\label{Fig4}}
\end{figure*}

As another example, we consider a tuneable frequency generator task \cite{Jae01}, where the input signal is a slowly varying step pulse signal that encodes a wave frequency but the desired output that ESN must produce is a sinusoidal wave of the frequency encoded by the step pulses. Figure~\ref{Fig4}a shows the output of ESN (the oscillations denoted by the blue line) to a typical step pulse signal (denoted by the red line). This result was obtained using software that accompanies Ref.~\cite{Jae01}. As we can see, the frequency of the sinusoidal signal generated by the ESN follows the amplitude of the step pulses. We will return to the discussion of this result in the following section.

\section{Physical Reservoir Computing and its Connection to Analogue Computers}
Soon after the introduction of ESN and LSM it was suggested that the reservoir (or liquid in the framework of LSM) does not necessarily need to be an artificial neural network that exhibits a nonlinear dynamical behaviour given by Eq.~(\ref{eq:RC1}) or its modifications \cite{Luk09, Gau21}. Indeed, it is plausible to assume that many real-life dynamical systems should be able to accept a certain input ${\bf u}_{n}$ and then produce observable neural activations ${\bf x}_{n}$ (liquid states in the framework of LSM) suitable for training an RC system to make a forecast. While the so-produced activation states may not necessarily fully describe the state of the RC system from the point of view of original ESN and LSM concepts, it has been shown that the replacement of an artificial neural network by a physical system that has similar nonlinear dynamical properties may help reproduce, to a significant extent, the essential functionality of a computational reservoir.

In fact, to date a number of physical systems, including spintronic devices \cite{Fur18, Rio19, Wat20}, quantum ensembles \cite{Fuj17}, electronic devices \cite{Cow05, Pen18, Nak21}, photonic systems \cite{Nak21, Sor20, Zen20, Sil21, Pan21, Che20_review, Raf20} and mechanical devices \cite{Nak21, Cou17}, have been used as a physical RC system that can operate similarly to the conventional algorithmic ESN and LSM. Yet, as noted in \cite{Luk09}, in some works on physical RC systems \cite{Nak15, Fer03, Jon07}, the concepts of ``reservoir'' and ``liquid'' were taken literally using a real reservoir filled with a physical liquid and considering disturbances on the surface of the liquid as neural activations. While, presumably, those attempts were motivated mostly by scientific research curiosity, they resulted in new knowledge and encouraging results that have motivated follow-up studies of nonlinear dynamical properties of fluids in the context of RC systems in particular and of analogue neuromorphic computing in general. 

Of course, the development of any physical RC system faces many fundamental and technical challenges. Indeed, whereas ESN and LSM are typically implemented as a computer program and, therefore, have a significant degree of flexibility that enables them to perform different kinds of forecasting tasks, the architecture of a physical RC system depends on the hardware or experimental setup that employs a specific nonlinear-dynamical behaviour for computational purposes. Subsequently, since many experimental setups consist of measurement equipment that is controlled by a computer, physical RC systems are often integrated and controlled by a digital computer or an auxiliary electronic circuit module to enable a higher degree of adjustability \cite{Sor19, Sor20, Wat20, Wat21, Raf20}.

On the one hand, such an approach restricts the ability of physical RC system to resolve different classes of problems since a particular physical system may not be able to accept certain types of input data. On the other hand, physical RC systems can be more efficient than digital computers  in resolving specific but very important practical tasks. In this sense, physical RC systems functionally resemble analogue computers that for many decades were more efficient than digital computers in predicting complex weather phenomena  \cite{Vee47, Hyn70}.

For example, Ishiguro's analogue computer \cite{Ishiguro, Miy20} employed a physical analogy between the electrical current and fluids \cite{Vee47} to model the height and flow of water in the North Sea. The sea was represented by an electronic grid consisting of several tens of nodes, where the flow of water depended on the difference in water height and on physical properties and civil engineering features of coastline constructions. Apart from the grid that modelled the sea, the analogue computer consisted of signal generators that synthesised time-varying inputs and of ancillary equipment such as an oscilloscope and a digital computer.

Thus, many features of the Ishiguro's analogue computer conceptually resemble the architecture of a physical RC system. Yet, not surprisingly, nowadays there is a resurge in the interest in analogue computers intended to be used in the field of hybrid computing \cite{Cow05, Ulm20}, biomedicine \cite{Ach16}, neuromorphic engineering \cite{Cra22, Ulm20}, quantum and molecular dynamics simulations \cite{Kop21} and education \cite{Ulm21}. The demand for such computers is dictated by difficult to overcome fundamental limitations of digital computers and specialised modelling software such as an insufficient capacity of algorithms to be effectively implemented on a parallel computer architecture \cite{Shi01}.   

It is noteworthy that Ishiguro's analogue computer was based on the fundamental physics of complex nonlinear processes that underpin sea-level changes, including material solitary waves \cite{Rem94} and tsunami waves \cite{Mad08}. Similar phenomena can also be observed in lakes, swimming pools, bathtubs and even in a bucket of water. This means that researchers may create a counterpart of Ishiguro's analogue computer using liquids contained in artificial reservoirs located in laboratory settings. Moreover, the same kind of waves can be controllably produced in microfluidic systems that have been actively investigated for the ability to perform certain digital computations \cite{Lee21}.  

Of course, liquids are much more difficult to control compared with electric signals propagating in electronic circuits. However, as shown, for example, in \cite{Mak18, Mak19}, liquids have one important advantage over solid-state technologies---they are highly nonlinear. Indeed, while achieving useful nonlinearity in electronic circuits requires a careful control of active circuit components such as transistors, in liquids the nonlinearity is always present naturally and it can be manipulated in a manner that is not attainable in solid-state systems \cite{Mak18, Mak19}. The same advantage of liquids applies when liquid-based technologies are compared with optical and photonic technologies, where nonlinear optics constitutes an independent field of research that aims to create novel nonlinear materials and techniques intended to overcome several fundamental physical limitations associated with nonlinear optical phenomena \cite{Mak19}.

To provide an intuitive example of how a liquid-based system might implement the standard ESN algorithm, let us consider a mm-sized bubble in a bulk of water (Figure~\ref{Fig4}c).  When the bubble is irradiated with a sound pressure wave, it harmonically oscillates, i.e.~it periodically inflates and deflates while maintaining a spherical shape \cite{Bre95, Lau10, Mak21, Mak22_bio}. The natural frequency of such oscillations is associated with the Minnaert resonance \cite{Min33} that occurs at the frequency given by the well-known formula:
\begin{eqnarray}
  f_{M} &=&\frac{\sqrt{3\kappa P_{a}}}{2\pi R_{0}\sqrt{\rho}}\,,
  \label{eq:Minnaert}
\end{eqnarray}
where the polytropic exponent of gas trapped inside the bubble is $\kappa=4/3$, $P_{a}$ is the pressure in the bulk of liquid outside the bubble, $R_0$ is the equilibrium radius of the bubble and $\rho$ is the density of the liquid. In this formula, the surface tension and the effect of viscosity of the liquid are negledted due to their insignificant impact on the oscillation frequency of mm-sized bubbles \cite{Mak21, Mak22_bio}.

Let us assume that the pressure in the liquid outside the bubble changes slowly in a stepwise manner and that the time interval between two consecutive step changes is long enough for the bubble to develop several periods of oscillations. For the sake of illustration, let us also neglect the decay of the oscillations at the time scale of a single step change in the pressure. The simulations are conducted for the step changes in the pressure that follow the pattern of the input signal of ESN in the frequency generator test in Figure~\ref{Fig4}a.

The result produced by this idealised model is presented Figure~\ref{Fig4}b. We can see that the first step change of the input signal leads to an increase in the hydrostatic pressure around the bubble, which, in turn, results in a decrease in the equilibrium radius of the bubble. In response to these changes, the bubble starts to harmonically oscillate around the new equilibrium radius with the frequency given by Eq.~(\ref{eq:Minnaert}) \cite{Bre95}. The hydrostatic pressure will change again due to the following step pressure change in the input signal, thereby forcing the bubble to oscillate around another equilibrium state with another frequency and so forth.

As a result of these processes, the frequency of the bubble oscillations in Figure~\ref{Fig4}b is changed similarly to the frequency of the ESN-driven tuneable frequency generator in Figure~\ref{Fig4}a, which indicates that an oscillating bubble can serve as an analogue version of the frequency generator implemented by the ESN algorithm (in fact, not only bubbles but also oscillation systems of other nature can be used for this purpose \cite{Cou17}).

A striking feature of the bubble-based frequency generator is its simplicity compared with ESN: while a bubble-based device can be made using technically simple and inexpensive equipment \cite{Mak21}, ESN requires a modern digital computer. On the other hand, a single bubble is suitable mostly for a sole task of generating harmonic waves with a frequency defined by the input. Nevertheless, as shown below, a cluster of interacting oscillating bubble can operate as a network capable of performing more complex prediction operations.

\section{Physical Reservoir Computing and Physics-informed Artificial Neural networks\label{sec:phys_informed}}
Recently, a family of advanced artificial neural network algorithms---physics-informed neural networks---was introduced, where models of networks strictly follow the laws of physics to make physically consistent forecasts \cite{Rai19, Doa20, Doa21}. From the mathematical point of view, to implement a physics-informed network, the standard ESN, LSM or similar algorithm needs to be combined with a mathematical model of a physical process that can be described using partial differential equations (PDEs) \cite{Rai19}. The so-designed RC systems have been shown to produce ML models that remain robust in the presence of noise or poorly defined data, which is often the case, for example, in biosensing \cite{Mak22_bio}. 

In a typical artificial neural network, the loss function such as the mean squared error (MSE) quantifies the difference between the expected real-life outcome and the forecast made by the ML model. Using the loss function one can obtain feedback that can be used to adjust the parameters of the model. Similarly to this approach, physics-informed algorithms exploit information that is collected both from measurements and from relevant mathematical models. In practice, this approach can be realised embedding known PDEs into the loss function of an artificial neural network.

\begin{figure*}[t]
 \includegraphics[width=0.69\textwidth]{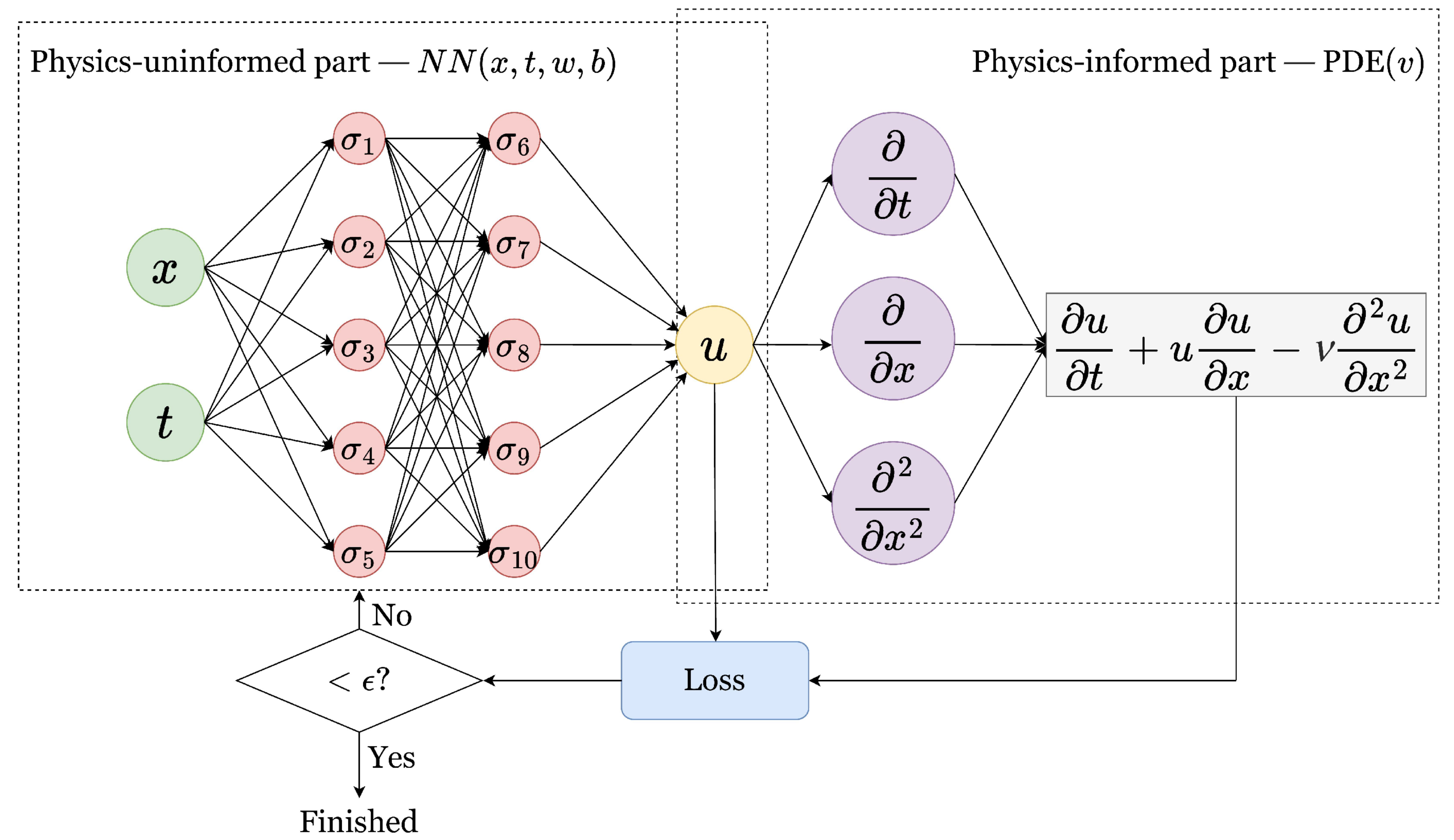}
 \caption{Block diagram of a physics-informed neural network that used a one-dimensional Burger's equation. Reproduced from Ref.~\cite{Mak22_Combs} under an open access Creative Common CC BY license and with permission of the authors.\label{Fig6}}
\end{figure*}

To provide a general example, Figure~\ref{Fig6} shows a block diagram of a physics-informed algorithm for solving a forward problem \cite{Hay98} using viscous Burger’s equation---an equation that often arises in the fields of nonlinear acoustics and fluid dynamics \cite{Mak19}. In a one-dimensional space defined by the coordinate $x$, for a given field $u(x,t)$ and the kinematic viscosity $\nu$ Burger’s equation reads:
\begin{eqnarray}
  \frac{\partial u}{\partial t} + u\frac{\partial u}{\partial x} = \nu \frac{\partial^2 u}{\partial x^2} \,.
  \label{eq:Burger}
\end{eqnarray}
In the block diagram in Figure~\ref{Fig6}, the conventional and physics-informed parts of the artificial neural network are denoted by the left and right dotted box, respectively. Both parts of the network are trained using the weights $w$, biases $b$ and a nonlinear activation function $\sigma$. A set of discrete time inputs $(x_i,t_i)$ is first fed into the network, then the Jacobian of the neural network’s output is computed using these inputs, and, finally, the residual of PDEs is computed and added as an extra term in the loss function. The entire network is trained by minimising the loss via a gradient-based optimisation method with a threshold parameter $\epsilon$.

It is plausible that the physics-informed neural network architecture shown in Figure~\ref{Fig6} as well as the physics-informed ESN algorithm \cite{Doa20} can be implemented using a physical system. A hybrid approach, where an artificial network is combined with a real-life physical system (as opposed to PDEs that describe the behaviour of that system), should also be possible. Apart from being of interest to researchers from the fundamental point of view, similarly to analogue computers discussed in the previous sections such a hybrid architecture may have certain advantages over digital systems.  

\section{Liquid-based Physical Reservoir Computers}
\subsection{Pioneering works}
We start our discussion with a brief overview of the results presented in the pioneering works on liquid-based physical RC system that implement the concept of LSM. According to some literature sources \cite{Fer03, Luk09}, the authors of those works took the idea of LSM literally and they were also motivated by the fact that waves on a liquid surface can be created using technically simple approaches \cite{Nat02}. While those results are well-known to experts in the field of neuromorphic computing and they have been discussed in several relevant literature sources \cite{Tan19, Nak20, Cuc22}, in general their analysis received a disproportionately little attention compared with discussions of solid-state RC system. Given this, here we discuss the main outcomes of those works with a focus on the aspects that, in our opinion, were not fully covered in the previous review articles. 

\begin{figure*}[t]
 \includegraphics[width=0.69\textwidth]{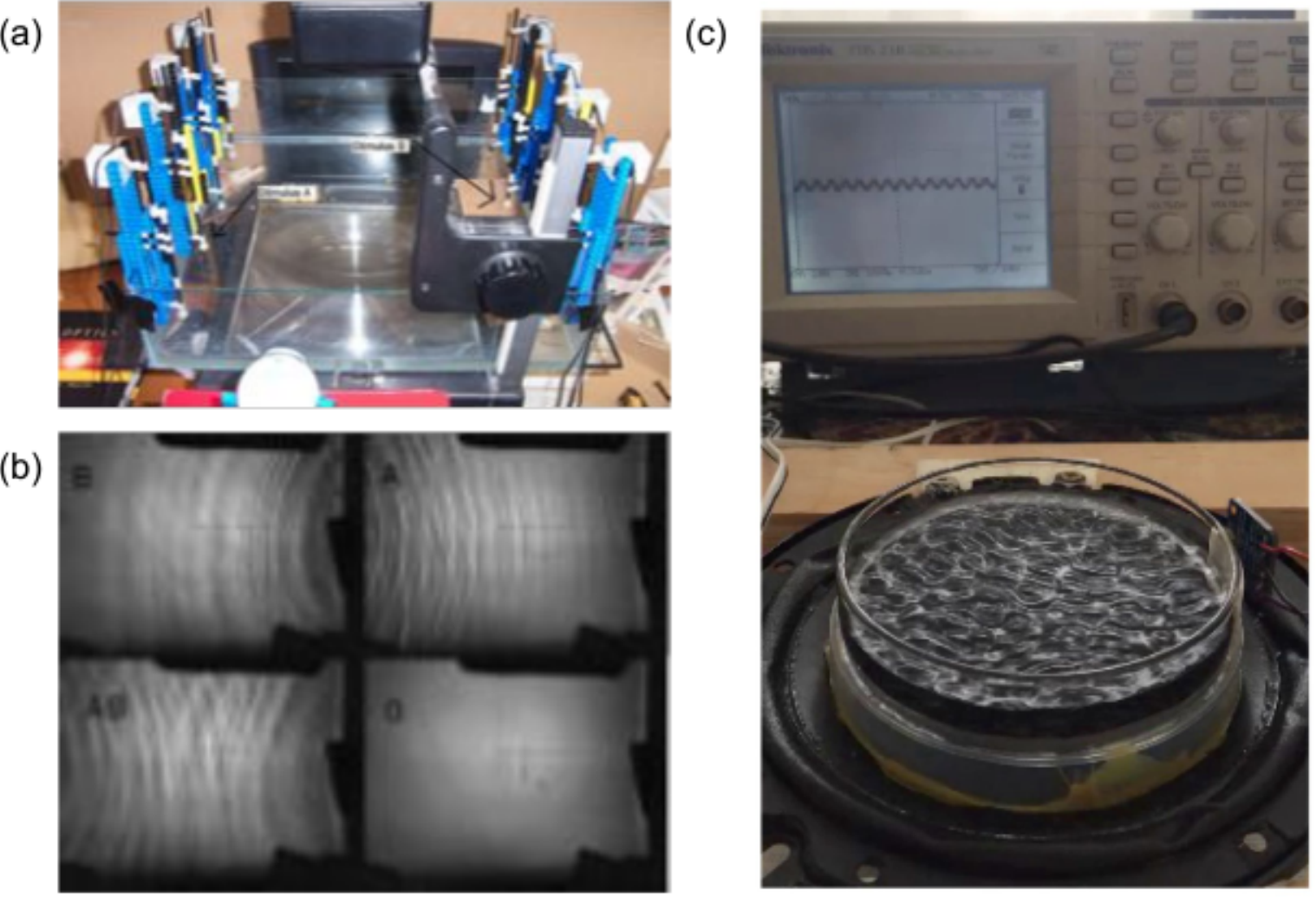}
 \caption{{\bf (a)}~The ``Liquid Brain'' physical implementation of LSM. {\bf (b)}~Photographs of the typical wave patterns for the XOR task. Top--Left: [0 1] (right motor on), Top--Right: [1 0] (left motor on), Bottom--Left: [1 1] (both motors on) and Bottom--Right: [0 0] (still water). Reproduced with permission from Ref.~\cite{Fer03}, Copyright (2003) by Springer Nature. {\bf (c)}~Photograph of Faraday waves on the surface of water contained in a Petri dish glued to a vibrating loudspeaker. Although the parts of the experimental setup shown on the photo were not used in RC-related experiments, the physics of Faraday waves observed here should be essentially similar to that in Ref.~\cite{Nak15}. Reproduced from Ref.~\cite{Mak22_Combs} under an open access Creative Common CC BY license and with permission of the authors.\label{Fig5}}
\end{figure*}

In the work \cite{Fer03}, the paradigm of LSM was implemented using a small optically-transparent tank filled with water (Figure~\ref{Fig5}a), and the resulting physical RC system was trained to undertake several challenging computational tasks, including a XOR test (Figure~\ref{Fig5}b). To create waves on the water surface, the tank was vibrated using several electric motors. The electric motors were so connected to each side of the tank that they enabled the creation of waves at eight different locations across the tank. The electric signals used to drive the motors were treated as the inputs of the RC system. A digital camera was used to register the interference pattern of the surface waves.  The resulting interference patterns were projected and recoded at a resolution of 320$\times$240\,pixels at 5\,frames per second rate (Figure~\ref{Fig5}b). Each frame was digitally processed to remove noise and to obtain 50 virtual neurons (see \cite{Wat21, Luk21} for a relevant discussion).

The authors of \cite{Fer03} supported the analysis of their results by a comparison of ``complexity of water'' with neural complexity \cite{Ton98}. It is well-established that a biological brain is high in complexity because of an intricate pattern of neural connectivity. Subsequently, the authors suggested that virtual neurons based on the dynamics of water waves might also exhibit a complex connectivity. To demonstrate the plausibility of that idea, they created a model of connectivity that revealed a staggered increase in connectivity due to an increase in the number of motors used to create water waves. The increase was interpreted to be due different effects of the individual motors on the water because the motors located in different places across the tank produced waves of different amplitude. It was also revealed that the RC system did not become chaotic when all motors were turned on, which is an important result because a dynamical system suitable for applications in RC system should not exhibit chaotic behaviour \cite{Luk09}. 

A conceptually similar approach was demonstrated in \cite{Nak15}, where vertical vibrations of a reservoir filled with a liquid resulted in the excitation of surface Faraday waves. Although the original article \cite{Nak15} is difficult to access, the results presented in it have been highlighted in some review articles (see, e.g.,~\cite{Nak20}). It is noteworthy that the dynamics of surface waves used in \cite{Nak15} is more complex than that of waves created in \cite{Fer03}, and, therefore, it may potentially be more useful for applications in RC systems. 

Indeed, classical nonlinear standing Faraday waves appear on the surface of a horizontally extended fluid in a vertically vibrating container \cite{Mak19_Faraday, Mak22_Combs} when the vibration amplitude of a harmonic signal with the frequency $f$ exceeds a critical value, which results in instability of the flat liquid surface and the formation of subharmonic surface waves oscillating at the frequency $f/2$ (Figure~\ref{Fig5}c). Speaking broadly, for example when the liquid forms a large pancake-shaped liquid drop, the formation of Faraday waves can result in an even more complex dynamical behaviour. In particular, the onset of Faraday waves in this case can be associated with the period-doubling bifurcation \cite{Mak19_Faraday}. The nonlinearity of such waves is so strong that one can observe a number of higher-order harmonics at the frequencies $nf$, where $n=2,3,\dots$. Clearly, when used judiciously, such a complex nonlinearity may be useful for applications in RC systems \cite{Jae01, Ber04}, especially when reservoirs with high memory capacity are required \cite{Gun08, Dam12, Kub21}.

\subsection{Bubble-based Physical Reservoir Computing}
Based on the discussion above, it is plausible to assume that a physical RC system should intrinsically be able to operate similarly to a physics-informed artificial network because a reservoir that operates using a real-life physical system is likely to produce activation states suitable for making a physically consistent forecast. This idea has been tested in \cite{Mak21_ESN}, where the nonlinear dynamics of a computational reservoir was replaced by that of a cluster of oscillating bubbles in water. In a previous work by the authors of \cite{Mak21_ESN}, it was demonstrated that mm-sized oscillating bubbles with randomly-chosen equilibrium radii and initial spatial positions produce acoustic signals that can be used to unambiguously measure the bubble dynamics in the cluster \cite{Mak21}. Therefore, it was suggested that a cluster consisting of $N_b$ bubbles can be used as a reservoir network of $N_b\times N_b$ random connections. In such a reservoir, the acoustic responses of individual oscillating bubbles serve as the neural activation states of algorithmic ESN \cite{Jae05, Luk09}. 

From the theoretical point of view, the implementation of this idea implies that Eq.~(\ref{eq:RC1}) is replaced by Rayleigh-Plesset equation that describes the nonlinear dynamics of spherical bubble oscillations \cite{Lau10, Mak22_bio} in a cluster consisting of $N_b$ bubbles that do not undergo translational motion (see \cite{Mak22_bio, Mak22_Combs} for a relevant discussion of translation motion):  
\begin{eqnarray} 
  R_p\ddot{R}_p
  &+&\frac{3}{2}{\dot{R}_p}^2 = \frac{1}{\rho}\left[P_p
      -P_\infty(t)\right]-P_{sp}\,,\label{eq:eq1}
\end{eqnarray}
where overdots denote differentiation with respect to time and for the $p$th bubble in the cluster
\begin{eqnarray}
  P_p = \left(P_0-P_v+\frac{2\sigma}{R_{p0}}\right)
      \left(\frac{R_{p0}}{R_p}\right)^{3\kappa} 
  -\frac{4\mu}{R_p}\dot{R}_p-\frac{2\sigma}{R_p}\,.\label{eq:eq2}
\end{eqnarray}
The term accounting for the pressure acting on the $p$th bubble due to scattering of the incoming pressure wave by the neighbouring bubbles in a cluster is given by 
\begin{equation}
  \label{eq:eq3}
  P_{sp}=\sum_{l=1, l\neq p}^{N_b}\dfrac{1}{d_{nl}}\left( R_l^2\ddot{R}_l
    + 2R_l{\dot{R}_l}^2 \right)\,,
\end{equation}
where $d_{pl}$ is the distance between the bubbles in the cluster and parameters $R_{p0}$ and $R_p(t)$ are the equilibrium and instantaneous radii of the $p$th bubble in the cluster, respectively. The pressure term $P_\infty(t)=P_0-P_v+\alpha_s u_s(t)$ represents the time-dependent pressure in the liquid far from the bubble, where $\alpha_s$ and $u_s(t)$ are the amplitude and temporal profile of the acoustic pressure wave used to drive the oscillations of the bubbles.

Once Rayleigh-Plesset equation has been solved, the acoustic power scattered by the $p$th bubble is calculated as \cite{Mak22_bio}
\begin{equation}
  \label{eq:eq4}
  P_{scat}(R_p,t) = \frac{\rho R_p}{h}\left(R_p\ddot{R}_p+2\dot{R}_p^2\right)\,, 
\end{equation}
where $h$ is the distance between the centre of the bubble and a spatial point in the acoustic far-field region \cite{Bre95, Lau10}. The resulting scattering pressure plays an important role in the integration of the bubble dynamics with the linear readout ESN algorithm. In Ref.~\cite{Mak21_ESN} this integration was achieved by sampling $P_{scat}(R_p,t)$ and $u_s(t)$ at equidistant time instances and obtaining their discrete analogs that were treated as the vectors of neural activations ${\bf x}_n$ and of input values ${\bf u}_n$ of standard ESN, respectively.
 \begin{figure*}
 \includegraphics[width=0.79\textwidth]{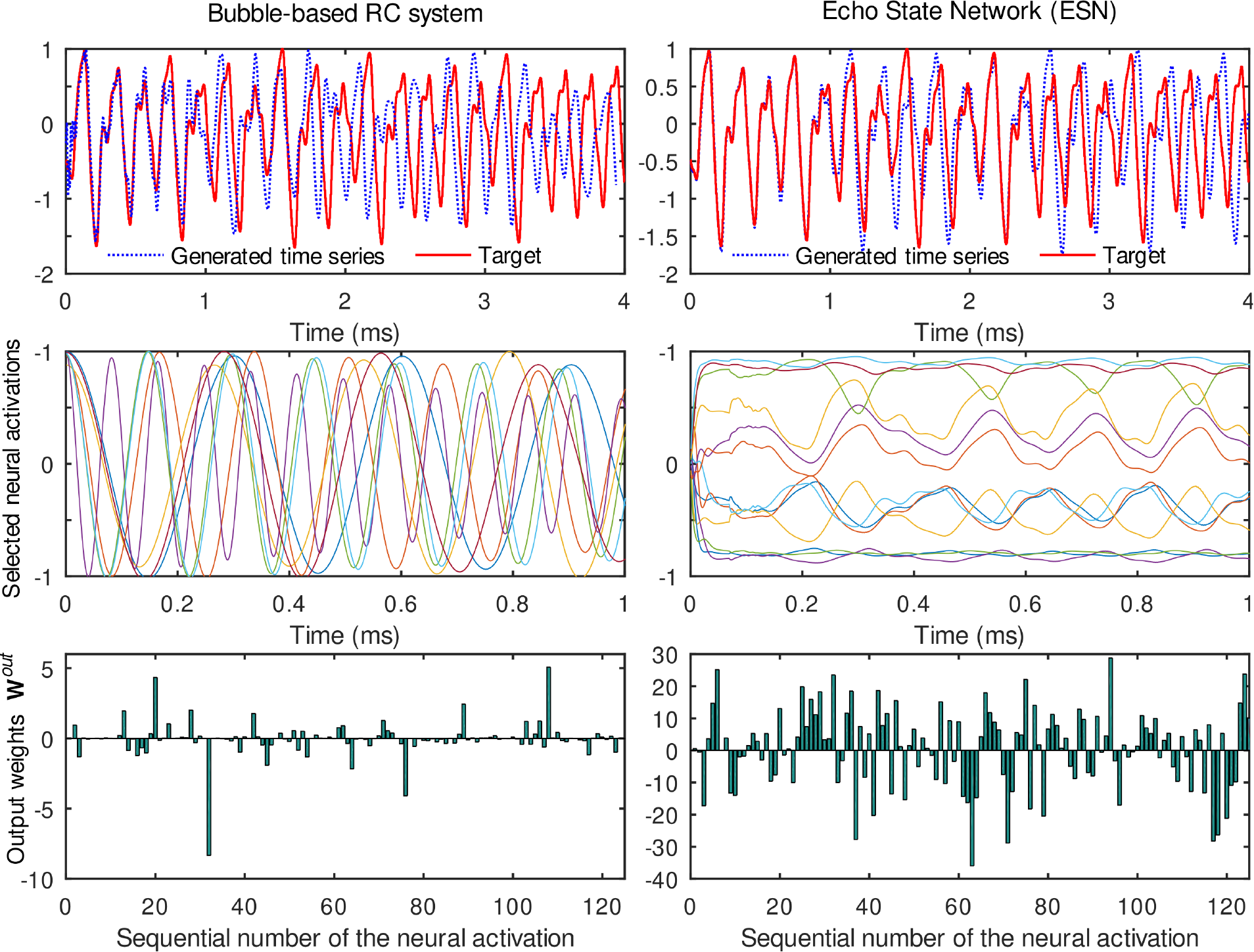}
 \caption{Performance of the bubble-based RC system (left column) and standard ESN (right column). Top row: target MGTS (the solid line) and its forecast made by the bubble-based RC system and ESN (the dotted line). Middle row: close-up view of selected neural activations. Bottom row: output weights ${\bf W}^{out}$ produced by the bubble-based RC system and ESN. Reproduced with permission from \cite{Mak21_ESN}, Copyright 2021 by the American Physical Society.\label{Fig7}}
\end{figure*}

In the bubble-based ESN, the signal $u_s(t)$ plays the role of the input signal that is used to train the network. The amplitude of $u_s(t)$ is chosen to be small ($\alpha_s = 0.1-1$\,kPa, i.e.~$\alpha_s \ll P_0$) so that both nonlinearity of the bubble oscillations and Bjerknes force of interaction between the bubbles remain weak. As a result, the echo state condition, which implies that dynamics of the neural activations ${\bf x}_n$ should be uniquely defined by any given input signal ${\bf u}_n$ \cite{Jae05}, becomes satisfied.  Moreover, with these model parameters, a cluster of mm-sized oscillating bubbles maintains its structural integrity for a period of time sufficient to train and exploit the RC system. The stability condition is important because the internal connections in the reservoir must not change during the training and exploitation of the network \cite{Luk12}.

Typical simulations conducted in \cite{Mak21_ESN} considered a cluster consisting of 125\,bubbles
with equilibrium radii randomly chosen in the 0.1 to 1\,mm range. The following physical parameters
corresponding to water at $20^\circ$\,C were also used: $\mu=10^{-3}$\,kg\,m/s, $\sigma=7.25\times10^{-2}$\,N/m, $\rho=10^3$\,kg/m$^3$, $P_v=2330$\,Pa, $P_0=10^5$\,Pa and $\kappa=4/3$.

The bubble-based RC system was tasked to forecast three different chaotic time series---Mackey-Glass time series (MGTS) \cite{Mac77}, Lorenz attractor \cite{Lor63} and R{\"o}ssler attractor \cite{Ros76}---and the results were compared with those produced by the standard algorithmic ESN that employed the same number of neurons, i.e.~there were 125\,neurons in ESN and 125\,bubbles in the bubble-based RC system.

For example, Figure~\ref{Fig7} compares the prediction made by the bubble-based RC system with that of standard ESN. We can see that in the time interval 0--1\,ms the bubble-based RC system correctly produces a relatively accurate forecast, which is evidenced by the calculated mean-square error (MSE) that is approximately $5\times10^{-2}$ for the bubble-based RC system and that is two orders of magnitude larger than MSE for ESN. Even though ESN performs better in this time interval, over the full test interval 0--2\,ms presented in Figure~\ref{Fig7}, MSE of both bubble-based RC system and ESN becomes approximately $5\times10^{-3}$, which indicates that the long-term forecast of the bubble-based RC system may be closer to the target MGTS than the forecast made by ESN.

It is noteworthy that the observed behaviour of ESN is widely regarded as a positive outcome for chaotic systems such as MGTS. Therefore, despite a somewhat inferior performance of the bubble-based RC system compared with ESN, the result in Figure~\ref{Fig7} clearly demonstrates the ability of a bubble-based physical RC system to predict chaotic time series.

Figure~\ref{Fig7} (middle row) shows neural activation states produced by the bubble-based RC system and ESN. The respective optimal weights ${\bf W}^{out}$ are shown in the bottom row of Figure~\ref{Fig7}. We can see that the behaviour of the activation states produced by ESN resembles the pattern of MGTS. However, the activation states of the bubble-based RC system are sinusoidal signals with the frequencies of oscillation of the bubbles in a cluster. This means that, while the
dynamics of the ESN is governed by the nonlinear operator $\tanh(\cdot)$ (see Eq.~(\ref{eq:RC1})), the dynamics of the bubble-based RC system is approximated by sinusoidal waves. The fact that the response of a nonlinear physical system can be represented as a superposition of sinusoidal
waves is well-established \cite{Kos93} and it has been used in the field of reservoir computing \cite{Sor19}. 

Finally in this section, we note the work \cite{Mak23_bubble}, where it has been demonstrated that a single oscillating bubble can operate as a physical RC system that has a substantial fading memory. In the cited paper, it has also been shown that such a single bubble can be used to produce musical outcomes, thereby serving as an AI system capable of producing creative works.
\begin{figure*}
 \includegraphics[width=0.89\textwidth]{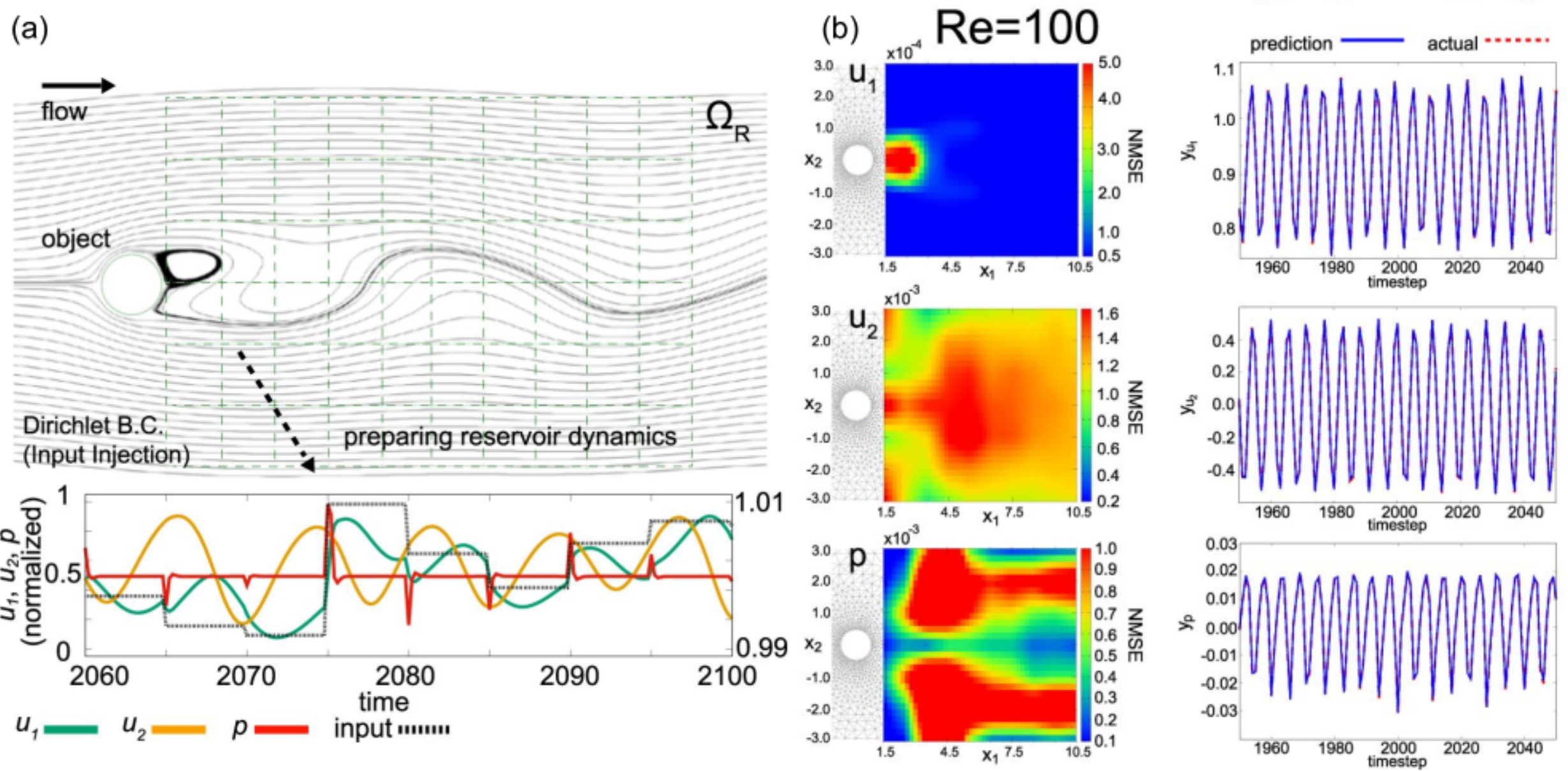}
 \caption{{\bf (a)}~Contours of a numerically calculated velocity distribution showing the vortex shedding. The direction of flow is indicated by the arrow in the top left corner. The mesh used to prepare the reservoir dynamics can be seen in the region past the circular cylinder that is denoted as ``object''. The bottom panel shows an example of the profile of the input flow alongside with the corresponding profiles of the velocity components and pressure obtained from one node of the mesh used to prepare the reservoir dynamics. {\bf (b)}~Result of the time series prediction task obtained at $\Rey=100$. The false colour maps shows NMSE for the target physical parameters of the fluid $u_1$, $u_2$ and $p$ in the two-dimensional computational domain covered by the mesh used to prepare the reservoir. The panels located next to each colour map show the prediction of the RC system made for a single discrete point of the computational domain (the expected actual signal is denoted by the dashed curve). Reproduced from \cite{Got21} under the terms of the Creative Commons Attribution 4.0 licence.\label{Fig8}}
\end{figure*}

\subsection{Reservoir Computing based on Vortices in Fluids}
As demonstrated previously in this article, different kinds of physical perturbations of the liquid surface can be employed to build a physical counterpart of the algorithmic LSM system. In the theoretical work \cite{Got21}, it has been suggested that vortices---regions in a fluid where the flow revolves around an axis line---can be used as such perturbations. Vortices are ubiquitous in nature and they can be observed, for example, as whirlpools, smoke rings and winds surrounding tropical cyclones and tornados.

Since the behaviour of vortices often exhibit a complex nonlinear dynamics that can be suitable for applications in reservoir computing, the authors of Ref.~\cite{Got21} carried out a rigorous numerical study to explore this opportunity. They chose a von K{\'a}rm{\'a}n vortex street that is well-known in the field of fluid dynamics as a repeating pattern of swirling vortices that are caused by a process known as vortex shedding \cite{Vor_book}. This phenomenon can be observed at a certain range of flow velocities that are defined Reynolds numbers ($\Rey$), typically when $\Rey>40$.

Very often, the formation of a vortex street is discussed using a system, where a fluid flows past a cylindrical object of infinite height \cite{Vor_book}. This classical two-dimensional system was used in \cite{Got21} as a computational reservoir (Figure~\ref{Fig8}a). As the input of the reservoir, the author used the modulation of the flow in front of the cylinder. To harvest the activation states of the reservoir, they discretised the space behind the cylinder with a fine rectangular mesh. Each node of the mesh contained the calculated values of velocity $\mathbf{u}={\left({u}_{1},{u}_{2}\right)}$ and pressure $p$ of fluid flow.

Since the dynamics of the so-constructed reservoir depends on the value of $\Rey$, several tests aimed to verify the performance of the reservoir were conducted for different values of $\Rey$ below and above the threshold of the formation of a von K{\'a}rm{\'a}n vortex street. Firstly, the memory capacity of the reservoir was tested employing a standard approach that is often used to establish the ability of an RC system to reproduce and nonlinearly process previous inputs using its current states \cite{Ber04, Wat20, Wat21} (however, Legendre polynomials were used as the input signal in \cite{Got21} instead of randomly generated square pulses used in \cite{Ber04, Wat20, Wat21}). In particular, it was shown that the effect of vortex shedding observed at $\Rey \approx 40$ resulted in the largest memory capacity of the system, which means that the reservoir should have the highest computational efficiency at this flow regime.

Secondly, the ability of the system to forecast chaotic time series was tested. Those tests revealed (Figure~\ref{Fig8}b) that the reservoir based on vortices can make accurate predictions of time series data also predicting physical parameters of physics flow similarly to physics-informed systems discussed in Sect.~\ref{sec:phys_informed}. The latter means that the reservoir dynamics may be prepared using some knowledge of dynamics of velocity and pressure of a fluid and then it can be used to produce complete spatio-temporal of velocity and pressure.
\begin{figure*}
 \includegraphics[width=0.79\textwidth]{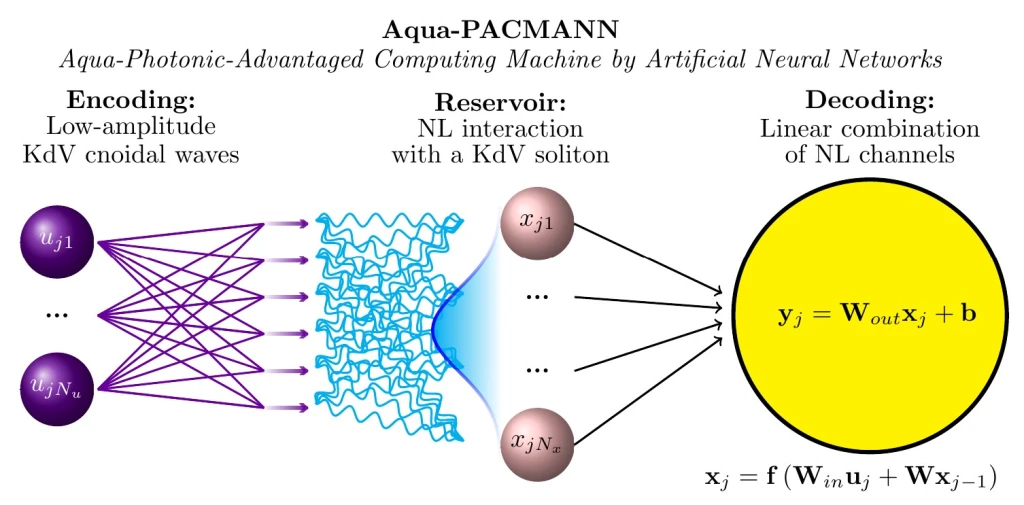}
 \caption{Illustration of Aqua-PACMANN physical RC system. The system consists of three layers. In the first layer, low-amplitude periodic waves represent the input information using the wave numbers and peaks amplitudes as data labels and numerical values, respectively. In the second layer that acts as the reservoir, the periodic waves propagate in shallow water and collide with a KdV soliton, resulting in nonlinear (NL) interaction processes. In the readout layer, the water wave amplitude is measured and sampled in time following the standard ESN algorithm, where a linear combination of NL information channels takes place. Reproduced from \cite{Mar23} under the terms of the Creative Commons Attribution 4.0 licence and with permission of the authors.\label{AquaPACMANN}}
\end{figure*}

\subsection{Physical Reservoir Computing based on Solitary Waves}
In this section we discuss physical RC systems based on solitary waves that are often observed in liquids. Several previous theoretical works \cite{Jun19, Zen20, Mar20, Sil21} have demonstrated the plausibility of physical RC systems based on solitary waves propagating in optical media \cite{Kiv03}. The reader interested in such RC systems is referred to the cited papers.  

Solitary waves maintain their shape and propagate with a constant velocity due to an interplay between nonlinear effects and dispersive processes in the medium \cite{Rem94, Kiv03}. Such waves have been of significant fundamental and applied research interest in optics \cite{Kiv03}, magnetism \cite{Sco05}, fluid dynamics \cite{Kor95},  electronics \cite{Xia09}, acoustics \cite{Per17} and biology \cite{Hei05, Gon14}. 

In the field of fluid dynamics, there have also been significant interest in solitary-like (SL) surface waves that originate from spatio-temporal evolution of falling liquid films \cite{Cha94, Kal12}. Due to their rich nonlinear dynamics and a relative technical simplicity of experimental implementation, SL waves have attracted significant attention \cite{Kap49, Yih63, Ben66, Shk71, Nak75, Ale85, Tri91, Gol94, Oro97, Ngu00, Thi04}. Furthermore, although SL waves share many physical features with the other kinds of solitary waves, they exhibit unique physical properties that are not observed in other physical systems \cite{Ker94, Gol94, Vla01}. For example, in contrast to two Korteweg-de Vries (KdV) solitary waves that represent waves on shallow water surfaces and that can pass through each other without significant change \cite{Kor95, Zab65}, two SL waves can merge, thus exhibiting an intriguing nonlinear dynamical behaviour \cite{Liu94}.
\begin{figure}
 \includegraphics[width=0.49\textwidth]{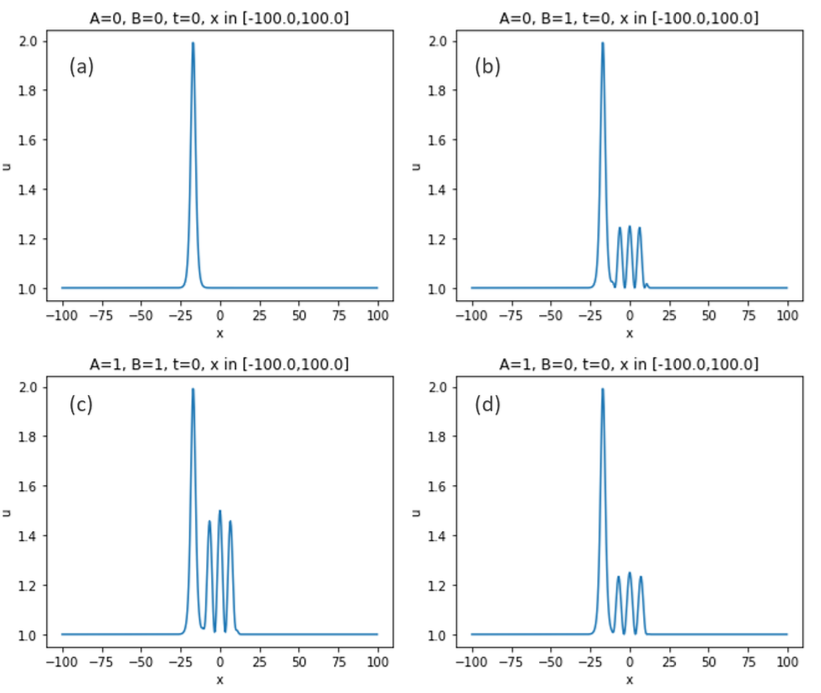}
 \caption{Low-amplitude waves that encode the input signals of the XNOR task. The titles of each panel denote the encoded Boolean variable couple. Reproduced from \cite{Mar23} under the terms of the Creative Commons Attribution 4.0 licence and with permission of the authors.\label{AquaPACMANN1}}
\end{figure}

\subsubsection{KdV Soliton-based Reservoir Computing}
In the paper \cite{Mar23}, a novel RC system that uses a superposition of KdV cnoidal waves as a method of information encoding and processing has been proposed and numerically validated. A cnoidal wave is an exact nonlinear periodic wave solution of the KdV equation \cite{Kor95}. Compared with a sinusoidal wave, a cnoidal wave is characterised by sharper crests and flatter troughs. Such waves can often be observed in nature, for example, as ocean and lake swell waves existing in shallow water \cite{Rem94}. 

The instantaneous wave velocity $u=u(x,t)$ of shallow water waves is described by the differential equation
\begin{eqnarray}
  \frac{\partial u}{\partial t} + u\frac{\partial u}{\partial x} + \beta \frac{\partial^3 u}{\partial x^3} = 0 \,,
  \label{eq:KdV}
\end{eqnarray}
where $\beta$ is a dispersion coefficient. In the case of travelling waves, the solution of Eq.~\ref{eq:KdV} has two limiting regimes known as low-amplitude waves and KdV solitons. In \cite{Mar23}, the former waves have been used to encode the input data but the latter ones underpined the operation of the physical reservoir, as shown in Figure~\ref{AquaPACMANN}. The authors called this system Aqua-Photonic-Advantaged Computing Machine by Artificial Neural Networks (Aqua--PACMANN) because they have envisioned that the detection of shallow water waves can be done using purely photonic devices such as a digital video camera. Specifically, it was suggested that Aqua--PACMANN could be implemented using a square-shape 10$\times$10\,cm container filled with a 1-cm-thick layer of water. The area of the container could be discretised with a spatial step of 1\,mm to obtain the nodes of the reservoir and harvest its states. It was also shown that a standard digital camera with the frame rate of 1000\,FPS would provide a suitable temporal resolution sufficient for training and exploitation of the computational reservoir.   

\begin{figure}
 \includegraphics[width=0.49\textwidth]{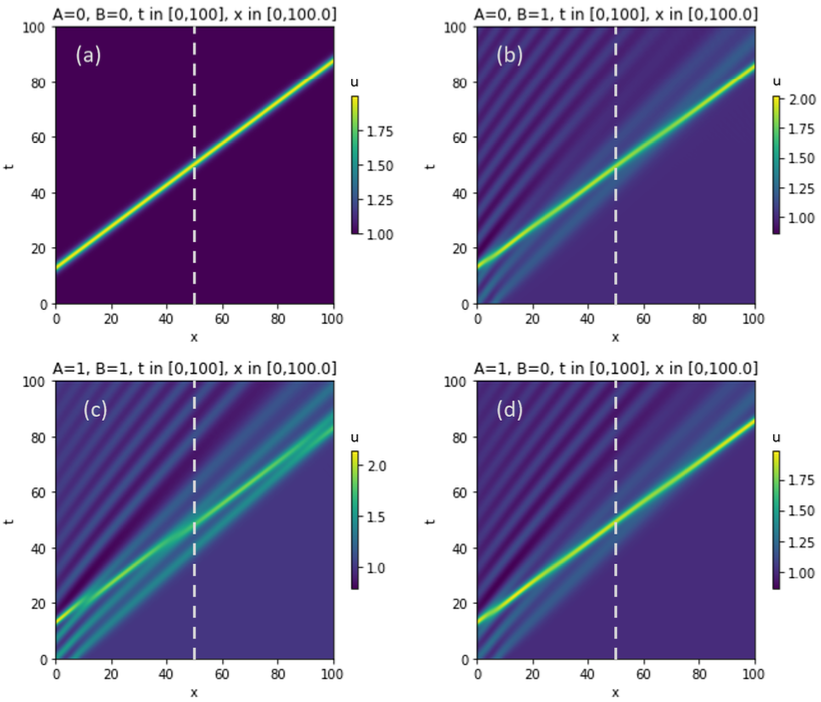}
 \caption{False colour maps showing the simulated evolution of KdV solitons with the initial states given by the XNOR truth table, as shown in Figure~\ref{AquaPACMANN1}. The titles of each panel denote the encoded Boolean variable couple. The vertical dashed line show the region, where the wave shapes are recorded and process to harvest the states of the reservoir. Reproduced from \cite{Mar23} under the terms of the Creative Commons Attribution 4.0 licence and with permission of the authors.\label{AquaPACMANN2}}
\end{figure}

The plausibility of Aqua--PACMANN to realise a logic gate was verified by tasking it with a XNOR test.  
Figure~\ref{AquaPACMANN1} shows the profiles of the waves used to encode the XNOR truth table and Figure~\ref{AquaPACMANN2} shows the result of the simulated spatio-temporal evolution of the input waves, which is the stage where the information processing takes place due to the nonlinear interaction. The reservoir states are harvested by measuring the wave profile in the centre of the reservoir that is denoted by the dashed line in Figure~\ref{AquaPACMANN2}. Using these data the weight matrix ${\bf W}^{out}$ is calculated, thereby concluding the training of the RC system and preparing it ready for the exploitation similarly to how the standard trained ESN system is used.   

\subsubsection{Reservoir Computing Using Solitary-like Surface Waves}
\begin{figure*}
 \includegraphics[width=0.69\textwidth]{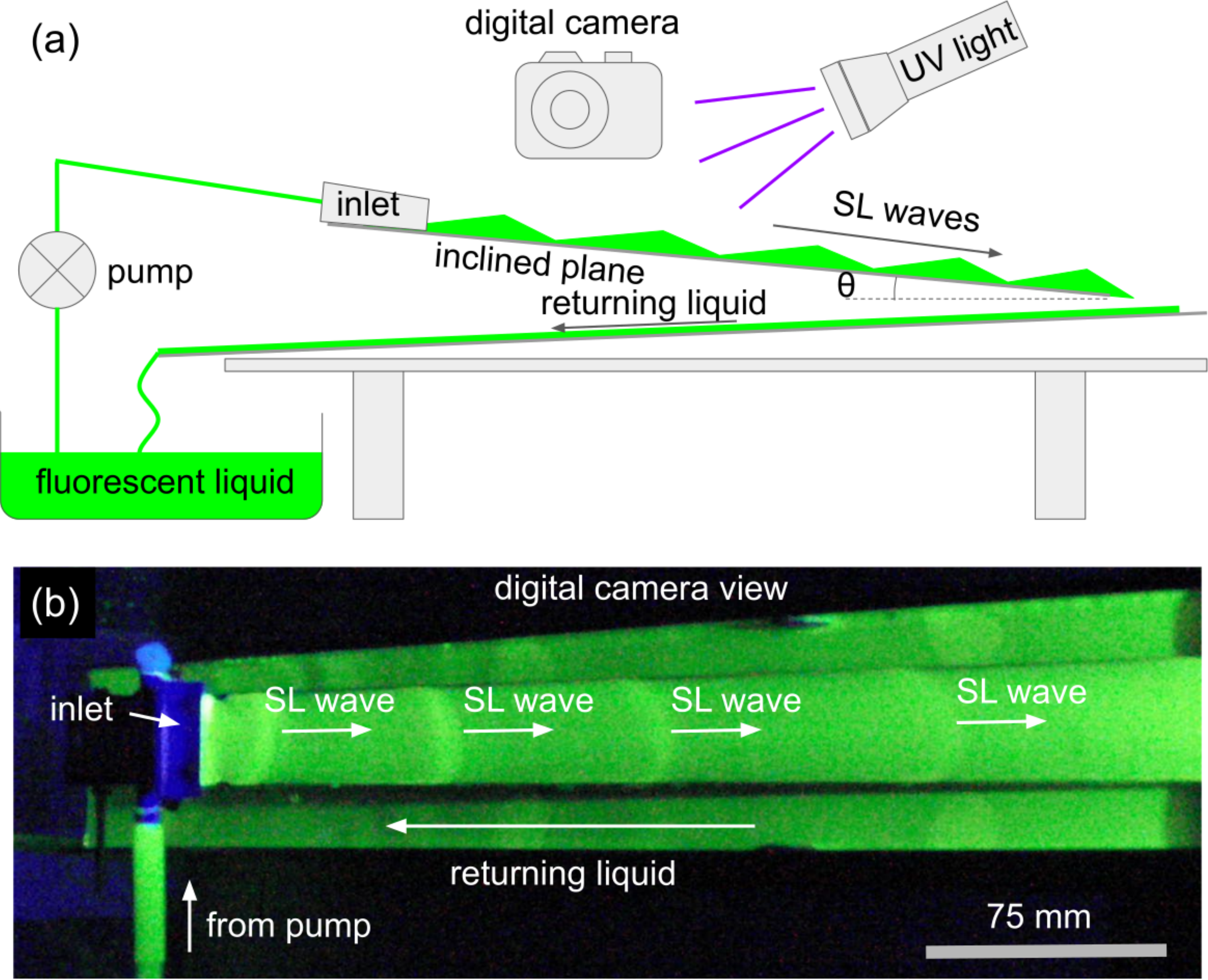}
 \caption{{\bf (a)}~Schematic reprsentation of the experimental setup of SL waves physical RC system. {\bf (b)}~Photograph of the experimental and SL waves taken by the digital camera. Reproduced with permission from \cite{Mak23_EPL}, Copyright 2023 by EPLA. \label{SLRC_review}}
\end{figure*}

A similar RC system that uses SL waves to encode the input information and to prepare the reservoir states has been proposed and experimentally verified in \cite{Mak23_EPL}. The experimental setup is shown in Figure~\ref{SLRC_review}a, where water flows over a metal plate that is inclined with the respect to the horizontal plane by the angle $\theta=3^{o}$. The flowing water forms a thin film with the thickness of approximately 1\,mm. The electric pump that was used to create the water flow is controlled by a microcontroller that modulates the water flow velocity to create SL waves (Figure~\ref{SLRC_review}b). Since the pump can reasonably follow low-frequency (0.1--5\,Hz) electric signals synthesised by the microcontroller, the SL waves can encode diverse input datasets, including complex signals such as MGTS.

To implement the standard ESN algorithm in the experimental setting depicted in Figure~\ref{SLRC_review}a, a digital camera was used and several possible detection mechanisms were tested. For example, the digital camera can used to detect just one image pixel located in the middle of the inclined plate, thus recording a time-dependent behaviour of SL waves that corresponds to just one neural activation of the reservoir. To produce additional neural activation,  a well-known sampling technique described in \cite{Wat20, Wat21} was applied. Alternatively, many image pixels located across the inclined plate can be used to create independent neurons. Both these approaches were tested and demonstrated an approximately the same efficiency. In all experiments, a fluorescent material was added to the liquid and a source of UV light was used to increase the visibility of SL waves.
            
The resulting SL wave RC system (SLRC) was tested using several standard benchmark tasks aimed to study the memory capacity of the reservoir and its ability to make free-running forecasts. When a trained RC systems is exploited in a free-running forecast regime, it uses its own outputs as the input data. It is noteworthy that this operating regime is, in general, challenging for algorithmic RC systems \cite{Luk09}. However, some physical RC systems have been able to successfully make such predictions even more efficiently than a standard ESN computer program \cite{Mak21_ESN}.

The experiments conducted in \cite{Mak23_EPL} revealed that SLRC has a very good memory capacity compared with the physical RC systems that use electronic, magnetic and optical devices as the reservoir \cite{Ber04, Nak21, Wat20, Wat21, Mat22}. This result is attributed to the unique property of SL waves to interact and merge, which means that the response of SLRC to one SL wave does not fully decay until the reservoir is presented with a series of following pulses.

Of course, to optimise this behaviour the amplitude and duration of SL waves and the spacing between them must be carefully controlled. SL waves also evolve as they move from the inlet in the downstream direction (Figure~\ref{SLRC_review}). All these characteristics of SL waves were numerically investigated using a hydrodynamic Shkadov model that is based on the boundary layer approximation of the Navier-Stokes equation \cite{Shkadov67, Shkadov68}. The modelling demonstrated that the optimal region for detection of SL waves is located in the middle of the inclined plate, which was the location used in the experiments in \cite{Mak23_EPL}.

\begin{figure}
 \includegraphics[width=0.49\textwidth]{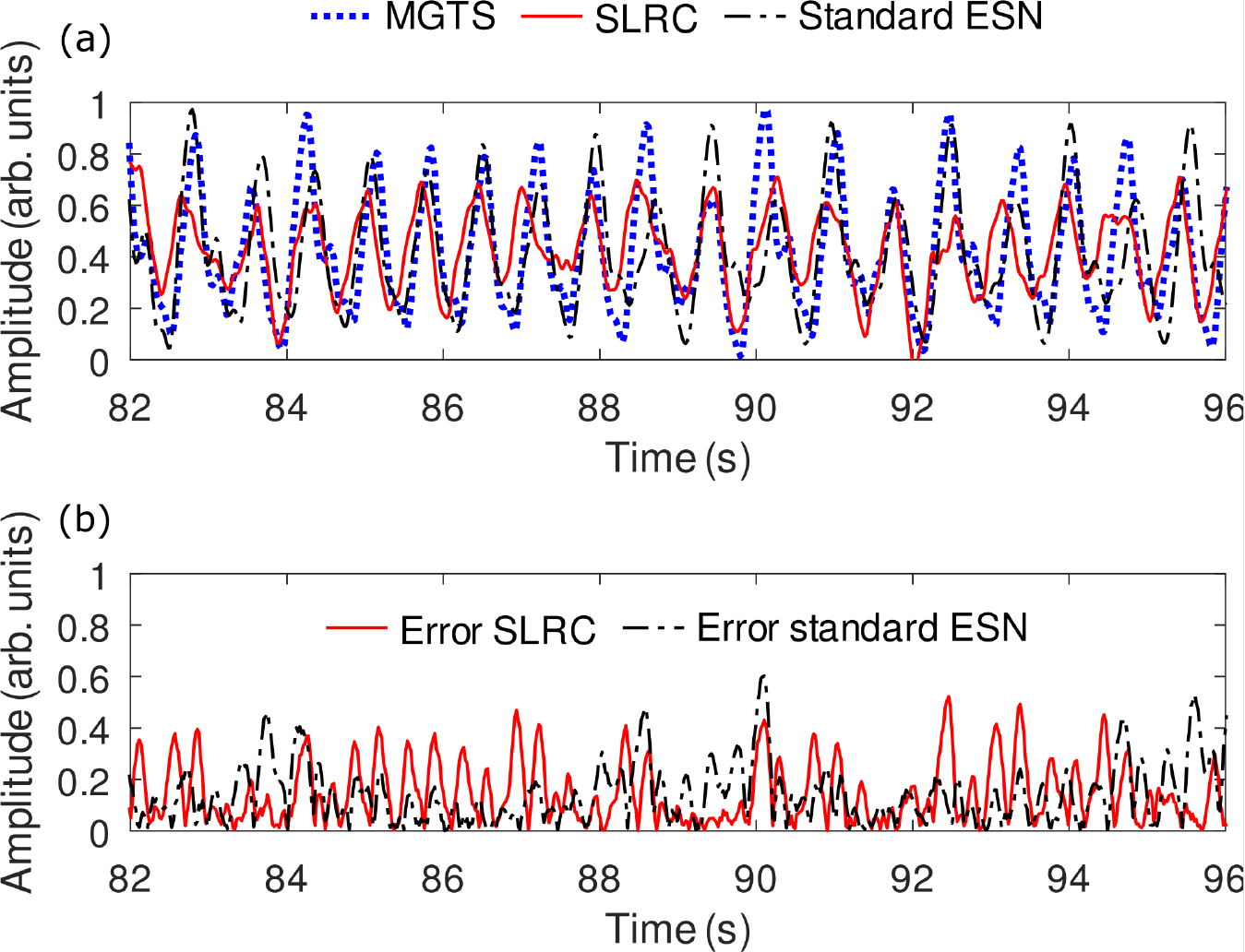}
 \caption{{\bf (a)}~Free-running forecast of MGTS (blue dotted curve) made by the SL wave RC system (labelled SLRC, red solid curve) and standard ESN (dash-dotted curve). Both RC systems used a reservoir consisting of 40\,neurons. {\bf (b)} Modulus of absolute error of SLRC (red curve) and ESN (dash-dotted curve). Reproduced with permission from \cite{Mak23_EPL}, Copyright 2023 by EPLA.\label{SLRC_MGTS}}
\end{figure}

The outcomes of the free-running forecast of MGTS are presented in Figure~\ref{SLRC_MGTS}a. We can see that SLRC can forecast MGTS with an accuracy of standard ESN that uses a reservoir with 40\,neurons. Due to current technical limitations, the number of neurons in SLRC could not be larger than 40. However, since the accuracy of the forecast made by SLRC (Figure~\ref{SLRC_MGTS}b) scales with the number of neurons similarly to the accuracy of standard ESN, it is expected that SLRC with about 200 neurons will produce a very accurate forecast of eat least the first five oscillations of MGTS.      

\section{Conclusions and Outlook}
The recent advances in the field of AI have changed the dynamics of the way we work and live. Nevertheless, AI systems remain unaccessible for most of the population in a large part due to their high complexity and cost. These problems are especially pressing in rural and remote areas, where the access to new technologies is often limited compared with the urban areas.

Physical RC system hold the promise to make some aspects of AI more accessible. In particular, the liquid-based RC architectures reviewed in this article can be employed as simple and inexpensive AI system implemented as a Microfluidic Processing Unit chip \cite{Lee21}. Such a chip may be integrated with hardware of a typical desktop digital computer, thereby converting the latter one into a much more computationally powerful but still financially affordable device that can replace a high-performace supercomputer while completing specific complex tasks.    

It is also noteworthy that high nonlinearity of water waves used in the reviewed physical RC systems is advantageous compared with nonlinear-dynamical phenomena observed in solid-state systems \cite{Mak19, Mak22}. Indeed, while significant theoretical and experimental effort have been made to understand nonlinear-optical effects and create optical devices based on them \cite{Mak19}, one can readily create water waves and exploit their nonlinear properties, which has been convincingly demonstrated in this article and the papers reviewed in it. The physical properties of water waves also align with the original paradigm of RC that exploits the concepts of reverberations \cite{Kir91} and of waves on the surface of an abstract liquid contained in a reservoir \cite{Maa02}.

Another considerable advantage of the liquid-based systems \cite{Tho16} is the possibility of complete elimination of electronic and photonic circuits that play an ancillary role in the operation of many physical RC systems but that often limit the speed of computations also increasing the complexity and cost of hardware \cite{Raf20}. Indeed, in the free-running forecast mode an RC system operates using its own past predictions as the input data \cite{Jae01, Luk09, Luk21}, which is achieved by means of a feedback loop that converts the output into input data at every discrete time step \cite{Luk21}. In a physical RC system implementing this approach, the feedback is realised using electronic or optical circuitry controlled by a computer \cite{Raf20, Sor20}. This feedback loop inevitably introduces a time delay that can be longer than the temporal scale of the dynamics of the reservoir \cite{Raf20}. Subsequently, the introduction of the feedback loop interrupts the natural dynamics of the reservoir. While in electronic and photonic feedback loops such an interference with the operation of the reservoir has been compensated using sophisticated experimental approaches \cite{Raf20}, a water wave feedback loop can be created naturally without introducing any delay \cite{Tho16, Mak23_EPL}.

Finally, the potential ability of water waves based RC systems to have a natural feedback enables them to be optimised and combined with more advanced algorithms such as the nonlinear vector autoregression (NVAR) machine \cite{Bol21, Gau21}. In Ref.~\cite{Mak23_EPL} its has been shown that a hybrid water based RC-NVAR system requires a lower training time compared with standard ESN without compromising the accuracy of forecasts. 

\acknowledgments{The author thanks Yaroslav Maksymov for preparing Figure~1 of this article and acknowledges multiple useful discussions with Dr Andrey Pototsky of Swinburne University of Technology.}

\bibliography{refs}

\end{document}